# Generation and transport of photoexcited electrons in single-crystal diamond

F. J. Heremans[1], G. D. Fuchs[1], C.F. Wang[1], R. Hanson[2], D. D. Awschalom[1]

*1. Center for Spintronics and Quantum Computation, University of California, Santa Barbara,*

*Santa Barbara, California 93106*

*2. Kavli Institute of Nanoscience, Delft University of Technology, Delft, The Netherlands*

**Abstract**

We report time-dependent photocurrent and transport measurements of sub-bandgap photoexcited carriers in nitrogen-rich (type Ib), single-crystal diamond. Transient carrier dynamics are characteristic of trapping conduction with long charge storage lifetimes of ~3 hours. By measuring the photoexcited Hall effect we confirm that the charge carriers are electrons and by varying the excitation energy we observe a strong turn-on in the photoconduction at ~1.9 eV. These findings shed new light on sub-bandgap states in nitrogen doped single-crystal diamond.

The combination of high thermal conductivity, large bandgap, and large dielectric breakdown make diamond attractive in optoelectronic, high-power and high-frequency applications. Previous photoconductivity studies have primarily focused on polycrystalline diamond grown using chemical vapor deposition (CVD) excited with bandgap energy photons[1,2]. While recent advances in diamond growth make single-crystal devices feasible, there remain fundamental questions regarding basic carrier transport properties, defect levels, and charge storage in these systems.

In this letter we investigate the room temperature photoconductivity of nitrogen-rich, single-crystal diamond in the presence of sub-bandgap energy illumination. Our samples are Sumitomo single-crystal type Ib diamonds grown with high-temperature, high-pressure methods and with a substitutional nitrogen density specified at $\sim 10^{19}$ cm$^{-3}$. Using standard lithography techniques, we fabricated Ti/Au gates on the sample with varying gap widths, shown in the inset of Fig.1(a). A voltage is applied across the two gates and the gap is optically excited by a 532 nm (2.3 eV) laser focused centrally between the two electrodes. In subsequent experiments, a tunable-wavelength laser is also used to vary the energy of the illumination. The resulting current flow is measured with a current pre-amplifier. Finally, samples in van der Pauw geometry are investigated to measure the photoexcited Hall effect.

Typical photocurrent-voltage (I-V) curves are shown in Fig. 1(a). We observe a scan-rate dependent hysteretic behavior that suggests charge is stored within the gap. First we establish whether these effects are due to bulk phenomena or to surface conduction, by measuring the current as we vary the laser focus of the 2.3 eV illumination through the gap in the electrodes. This measurement was taken for 10 μm, 20 μm, and 50 μm gap widths, Fig. 1(b). The two maxima correspond to focal positions where the illuminated volume bridges the gap with maximum intensity (see Fig. 1(b), inset). At positive focus (out of the sample), the current is larger than at negative focus (into the sample) due to the larger illuminated volume. Moreover, the positions of the maxima track the width of each gap. These results confirm that the photocurrent is proportional to the illuminated volume, which rules out a dominant surface effect. In order to further suppress surface conduction, recently understood to result from a hydrogen terminated diamond surface[3,4], we either annealed the sample or purged the surface with dry nitrogen during the measurement.

To study the storage of charge within the gap, we use a charge/discharge sequence, shown in Fig.2(a). Initially, we apply an electric field in the presence of the optical excitation, which induces a photocurrent. The laser is then blocked and, after a constant 10 s delay, the external electric field is removed. This results in a transient current that flows with the opposite sign as the photocurrent. We attribute this to a partial relaxation of the space-charge electric field[5] that opposes the externally applied field (inset of Fig. 2(a)) but is difficult to quantitatively model in this geometry and beyond the scope of this work. After a given delay (dark time), we resume the laser illumination, which causes an additional transient current. This discharge also has the opposite polarity as the initial photocurrent, indicating it is also caused by the space-charge field within the gap.

The current measured versus time in the charging and discharging curves are well fit by a stretched exponential form,

$$I = I_0 \exp\left[-\left(\frac{t - t_0}{\tau}\right)^P\right] \quad (1)$$

shown in Fig. 2(b), where $I_0$ is the amplitude, $t_0$ is the initial time, $\tau$ is the characteristic time constant, and $P$ is a power constant. Fitting the data to Eqn. (1) yields consistent values for the power exponent $P$, for individual gap widths, independent of delay time and applied voltage. $P$ varies from 0.37 for the 10 μm gap to 0.54 for 50 μm, which is consistent with values of stretched-exponential transient currents in illuminated CVD diamond[6,7]. The stretched exponential was theoretically predicted for conduction by carriers originating from either bands of traps[8] or discrete trap states[9] within the gap. The saturation of current in the initial charging sequence shown in Fig. 2(a) can be explained by these trap states within the gap being fully populated. Such mechanisms are consistent with both the observed space-charge effect and charge storage effect shown in Fig. 2. As expected, the amplitude $I_0$ decreases with increasing delay times as the charge slowly dissipates out of the traps. The time constant $\tau$, in general, decreases with longer dark times as more traps vacate, facilitating photoconduction.

Integrating the discharge curve yields the amount of charge stored for a given dark time, shown in Fig.2(c). The characteristic time for charge stored in the 10 μm gap is 2.5 hours and 3.3 hours for the 50 μm gap. However, there is still a substantial charge observed after 14 hours due to the long tail in the stretched exponential. Fig. 2(c) also shows an expected result: larger gaps have a larger illuminated volume and hence a greater number of traps that become filled. We observe that substantially more charge is stored in the 50μm gap than in the 10μm gap, which further confirms this picture. When we vary the laser power across a given electrode gap, we find the total charge stored is independent of the laser power provided the charging time is sufficiently long to fully populate the traps (inset of Fig.2(c)).

The observations of stored charge led us to suspect the physical origin of the trap states are substitutional nitrogen defects (also known as P1 centers), which are dominant in Ib diamond. Substitutional nitrogen has previously been observed to form a trap state and is expected to be an electron donor[10]. In order to test this hypothesis, we fabricated a sample with electrodes in the van der Pauw geometry and measured the photoexcited Hall voltage. The measurements demonstrate a negative Hall voltage corresponding to n-type conduction, confirming that the photoexcited carriers are electrons.

We also performed excitation-energy dependent photoconductivity measurements to elucidate the origin of the trap states. A tunable, pulsed laser with excitation energies between 1.68 eV and 2.19 eV was used to excite photocurrents across a 70 μm gap. As expected, the photocurrent scales linearly with the laser power for all pump energies, but with a slope that depends strongly on the excitation for energies above 1.9 eV. Fig. 3(a) shows representative plots of laser power versus photocurrent. To further understand this cross-over in behavior as a function of excitation energy, we set the laser power at a constant 10 mW, and measured the photoexcited I-V characteristic at high voltages, where there is a linear response (30-70 V), shown in Fig. 3(b). We minimize the hysteretic behavior and charge stored within the gap by implementing slow scan rates and long wait times. Nevertheless, history-dependant residual charge forms the largest systematic error.

From these data, we determined the differential conductance (dI/dV) which is plotted in Fig. 3(c). There is a strong turn-on of dI/dV around 1.9 eV, suggesting an impurity level with states that extend to 1.9 eV below the conduction band. A slight peak in the differential conductance, on the order of the signal-to-noise ratio, is also observed around 1.75 eV. The observed 1.9 eV turn-on and 1.75 eV feature are found in the same energy regime as the proposed substitutional nitrogen donor defect ionization energy around 1.7 eV below the conduction band[11]. Our findings closely agree with a previously observed thermally stimulated electron trap level located at 1.9 eV below the conduction band[12], as well as a 1.945 eV vibronic band[13] in type Ib diamond.

The origin of the 1.9 eV turn-on observed in our experiments has yet to be confirmed. However, previous experiments, along with nitrogen being the dominant defect in type Ib diamond, suggest the trap states are due to the presence of the substitutional nitrogen defects (P1 centers). Electrons are photoexcited out of these traps and move along the electric field until they fall into vacant traps. The net conduction of electrons moving from one electrode to the other across the gap via this trapping/re-trapping mechanism constitutes the measured photocurrent. Whether these nitrogen trap states are discrete traps or a band of traps is still uncertain. The stretched exponential function seen in the time-dependent current transport suggests it can be either[8,9]. The crystalline nature of diamond and the sharpness of the 1.9 eV turn-on, however, are consistent with a discrete trap level.

In conclusion, the bulk photoexcited transport demonstrates a trapping/re-trapping conduction mitigated by space-charge. In addition, the trap states have a long lifetime, with a characteristic decay time of ~3 hours. Energy-dependence measurements show the trap states lie 1.9 eV below the conduction band, which is consistent with substitutional nitrogen defects in the diamond[12,13].

We thank C. G. Van de Walle for discussions, and R. J. Epstein for preliminary measurements, and AFOSR and ARO for their financial support. A portion of this work was done in the UCSB nanofabrication facility, part of the NSF funded NNIN network.

## Figure Captions

FIG. 1. (a) I-V measurements showing a hysteretic loop that widens as the voltage scan rate increases, suggesting charge is being stored within the 10 μm gap. [inset: Ti/Au gates (light gray) deposited on a type Ib diamond (dark gray) with gaps varying from 10 – 100 μm] (b) Normalized photocurrent varies with the depth of focus. Positive values in focus correspond to values out from the sample surface (0 μm).

FIG. 2. (a) The charge/discharge sequence. Initially both the optical excitation and applied field are on. The laser is then blocked, followed by the removal of the voltage bias. A dark time is varied and then the laser is unblocked and the current is measured. [inset: schematic of the space-charge electric field within the sample at during the charge/discharge sequence] (b) The discharge current transients decays as a stretched exponential function (solid lines show fit). Various dark times are shown and the decay time and amplitudes decreases as the dark time increases. Data shown is across a 10 μm gap with a 4 V applied bias. (c) The stored charge is calculated by integrating an individual discharge curve in (b). As expected the 50 μm gap (blue) stores more charge than the 10 μm gap (red), as there are more traps within the region. The stored charge fit also follows the stretched exponential form with a similar value of $P$. [inset: stored charge as a function of optical density]

FIG. 3. (a) Photoexcited current as a function of laser power at a 50 V gate bias across a 70 μm gap shown for four different excitation energies. (b) I-V measure taken as for four excitation energies (solid lines show linear fit of differential photoconductance). (c) The differential photoconductance, dI/dV, as a function of excitation energy, showing a turn-on around 1.9 eV

(arrow). [inset: a schematic energy band diagram of the observed effect. CB, conduction band, VB, valence band]

Figure 1

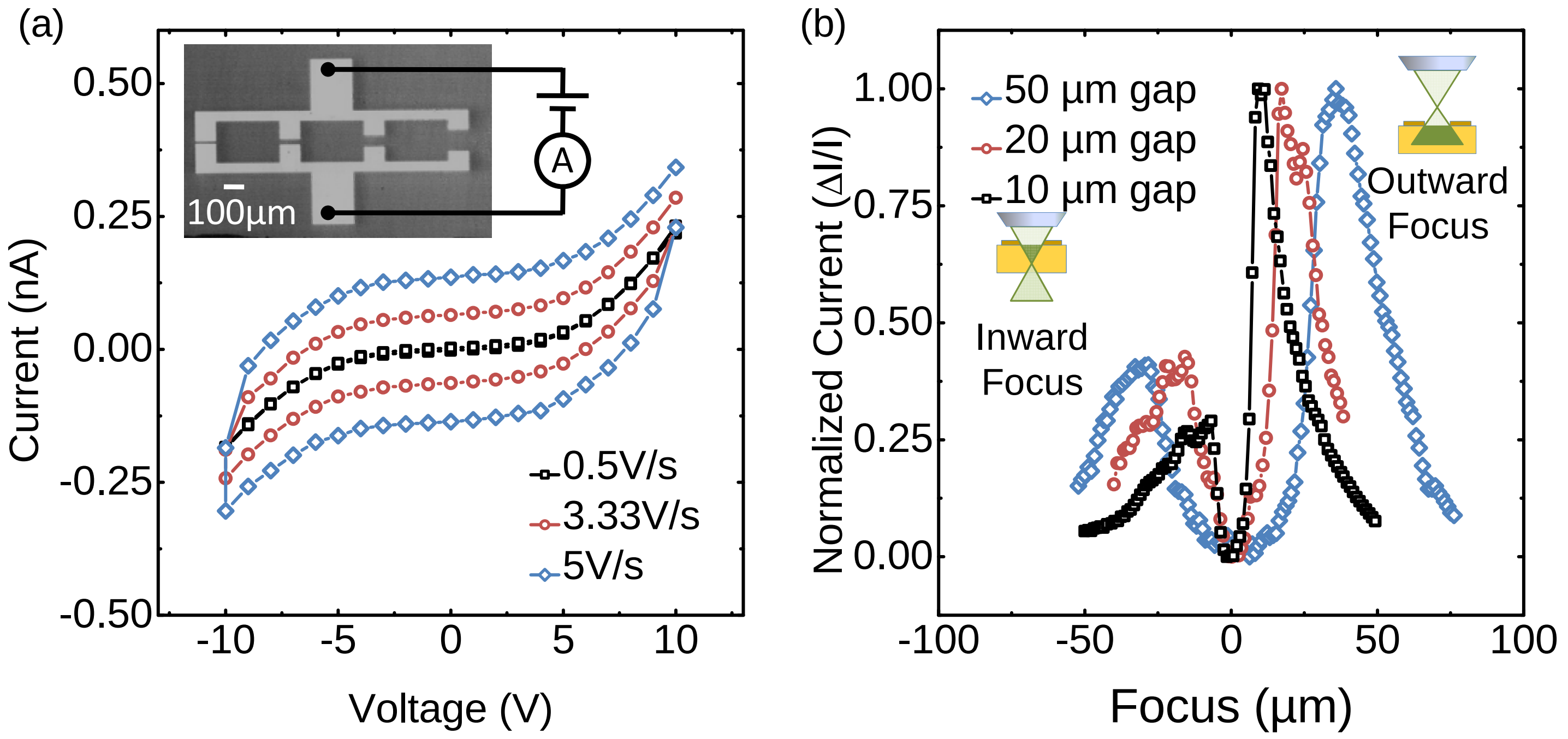

Figure 2

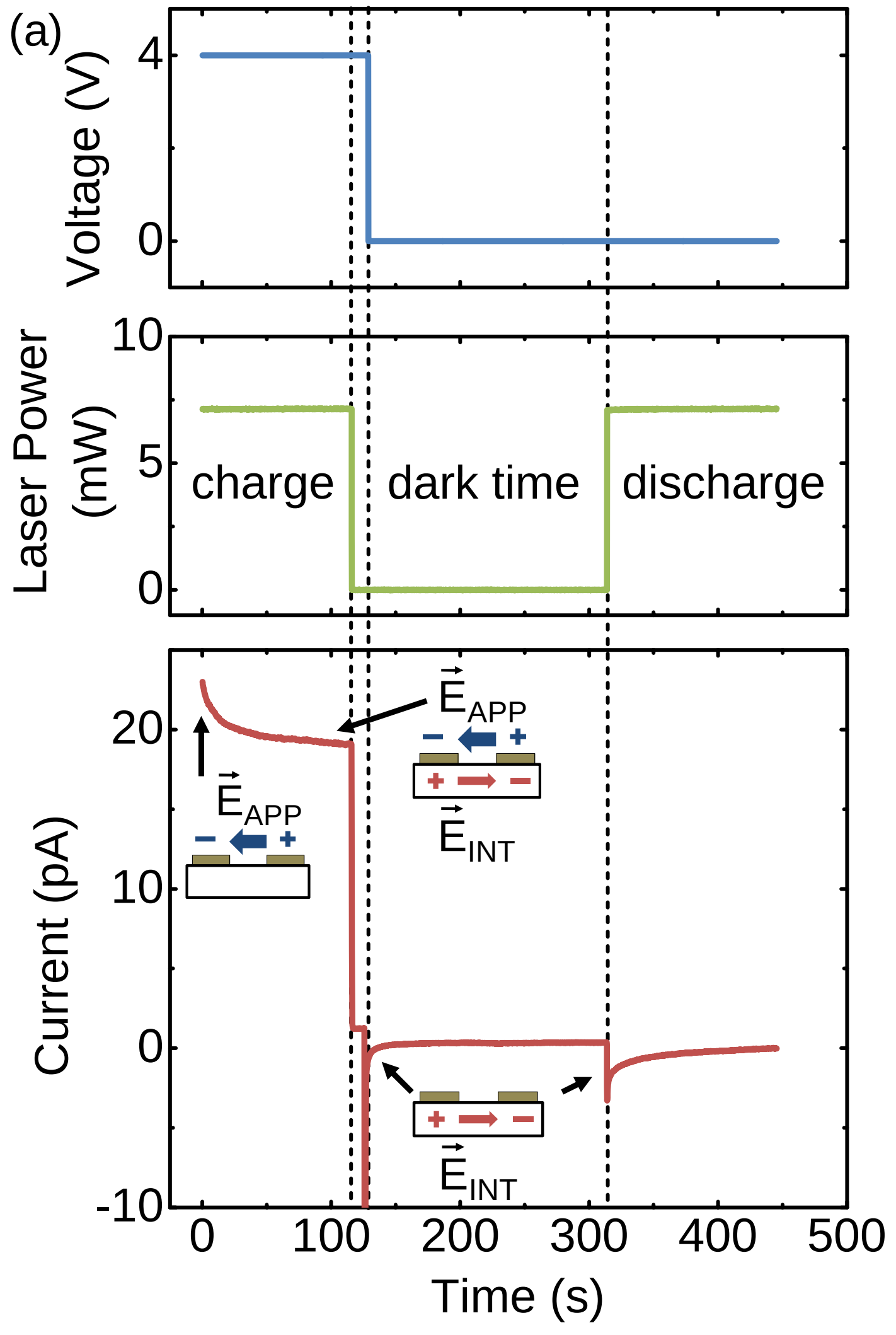

Figure 3

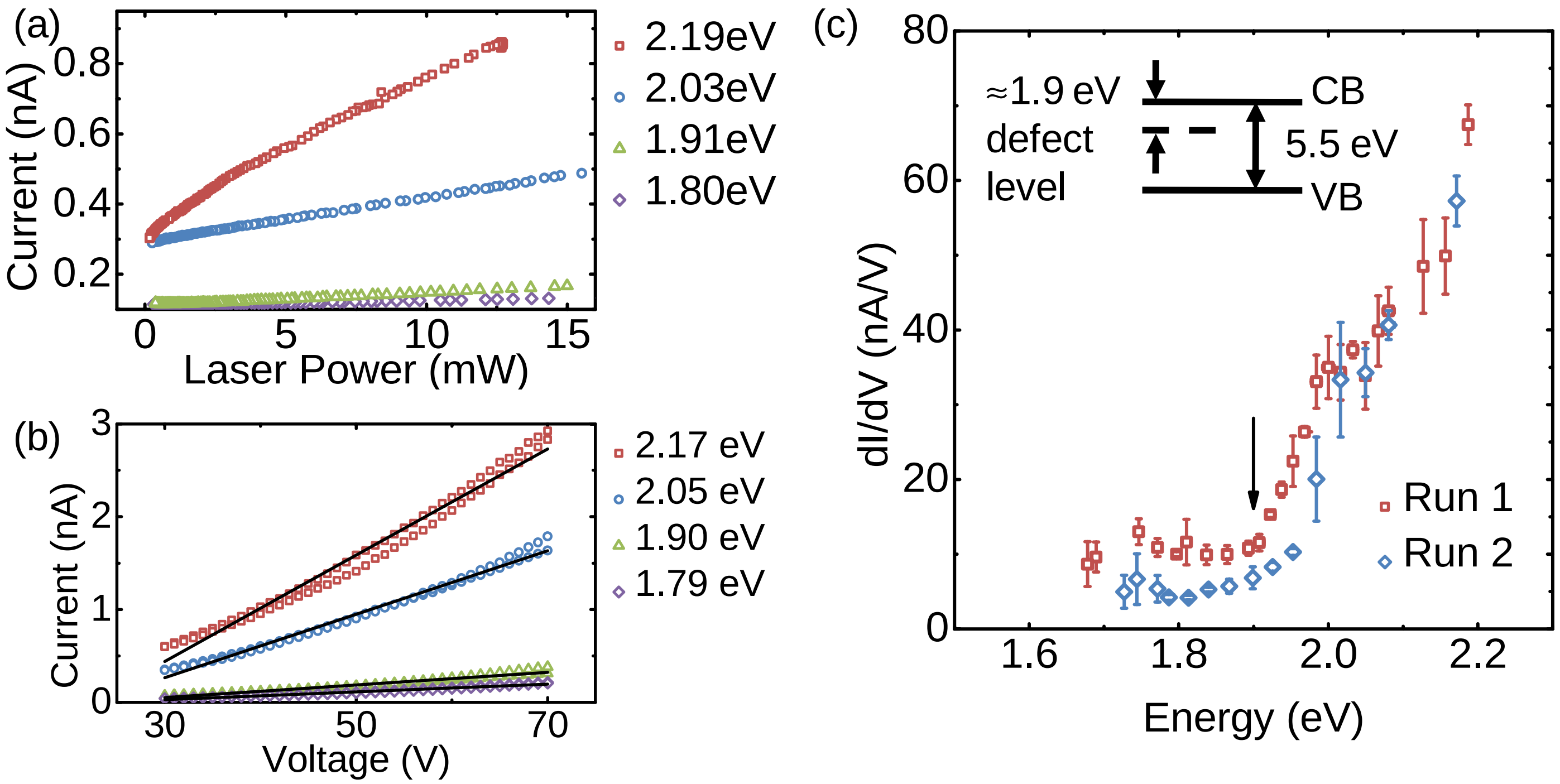